\documentclass{ws-procs9x6}

\begin{document}

\title{Rescattering effects and two-step process\\ 
in kaon photoproduction on the deuteron}
     
\author{A. Salam\footnote{\uppercase{P}ermanent address :  
\uppercase{D}epartemen \uppercase{F}isika, \uppercase{FMIPA},  
\uppercase{U}niversitas \uppercase{I}ndonesia, \uppercase{D}epok 16424, \uppercase{I}ndonesia.}~
 and K. Miyagawa}

\address{Simulation Science Center, 
Okayama University of Science, \\
1-1 Ridai-cho, Okayama 700-0005, Japan}

\author{H. Arenh\"ovel}
\address{Institut f\"ur Kernphysik, Universit\"at Mainz,
D-55099 Mainz, Germany}

\author{T. Mart}
\address{Departemen Fisika, FMIPA, Universitas Indonesia, 
Depok 16424, Indonesia}  

\author{C. Bennhold}
\address{Center for Nuclear Studies, Department of Physics,\\ 
The George Washington University, Washington, D.C. 20052, USA}

\author{W. Gl\"ockle}
\address{Institut f\"ur Theoretische Physik II, Ruhr-Universit\"at Bochum,\\
D-44780 Bochum, Germany}

\maketitle

\abstracts{
Kaon photoproduction on the deuteron is investigated 
by considering $YN$ and $KN$ rescattering 
and the two-step process 
$\gamma d \rightarrow \pi NN \rightarrow KYN$. 
A strong enhancement in the total cross section 
is found from the two-step process. 
$YN$ rescattering has remarkable effects 
in the inclusive and exclusive cross section, 
while the effect of $KN$ rescattering is much less important.
}

\section{Introduction}
\label{sec-introduction}

Kaon photoproduction on the deuteron has been investigated by several people. 
Li {\it et al.}\cite{LiW92} have extracted 
the elementary cross section from the deuteron target 
in the study of neutron channels. 
In a recent paper Yamamura {\it et al.}\cite{YaM99} 
have calculated the $YN$ rescattering 
for the $K^+$ channels by using the Nijmegen
$YN$ potential\cite{MaR89}. They found sizeable effects in 
the inclusive cross sections from the $YN$ rescattering. 
This work is extended by considering 
the two-step process $\gamma d \rightarrow \pi NN \rightarrow KYN$ 
and the $KN$ rescattering\cite{Sal03}. 
Other recent calculations also investigated the $YN$ rescattering\cite{Ker01} 
and the pion mediated process in lowest order\cite{Max04}.

This paper presents the calculation of $K^+$ and 
$K^0$ photoproduction on the deuteron by considering 
$YN$ and $KN$ rescattering and the two-step process. 
The formulations are shown in Section~\ref{sec-formulations}, 
the results in Sect.~\ref{sec-results}, 
and the conclusions in Sect.~\ref{sec-conclusions}. 

\section{Some Formulations}
\label{sec-formulations}

\begin{figure}[htbp]
\centerline{\epsfxsize=4.1in\epsfbox{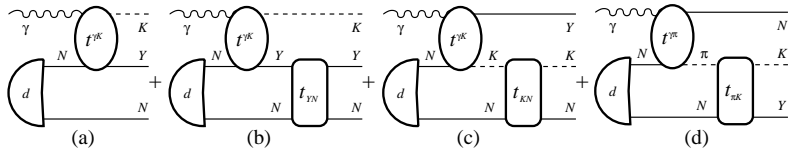}}   
\caption{Kaon photoproduction on the deuteron. 
Diagram (a) is impulse approximation (IA), 
(b) and (c) are $YN$ and $KN$ rescattering, respectively, and 
(d) is the two-step process ($\pi K$-process for short). 
\label{fig-gdkyn-diagram}}
\end{figure}

In the deuteron rest frame the exclusive cross section is given by
\begin{eqnarray}
\frac{d\sigma}{dp_{K}d\Omega_{K}d\hat q_{Y}} &=&  
\frac{m_{Y}m_{N}\vert\vec p_{K}\vert^2\vert\vec q_{Y}\vert}
{4(2\pi)^2E_{\gamma}E_{K}W} 
\frac{1}{6} \sum_{\mu_{Y}\mu_{N}\mu_{d}\lambda} 
\left\vert \sqrt{2}\langle\Psi^{(-)}_{\mu_{Y}\mu_{N}}
\vert t^{\gamma K}_{\lambda}\vert
\Psi_{\mu_{d}}\rangle\right\vert^2
\label{eq-gdkyn-dsigma-inclusive}
\end{eqnarray}
where $\mu_{Y}$, $\mu_{N}$, $\mu_{d}$, and $\lambda$ denote 
the spin projections of hyperon, nucleon, deuteron, and the photon 
polarization, respectively, $W^2=(P_{d}+p_{\gamma}-p_{K})^2$, 
and $\sqrt{2}$ comes from the proper antisymmetrization. 
The amplitude is approximated by the diagram in Figure~\ref{fig-gdkyn-diagram}. 

The impulse term and $YN$ rescattering (diagram (a) and (b), respectively) 
are calculated simultaneusly as
\begin{eqnarray}
T_{IA+YN} &=& t^{\gamma K} + t_{YN}\, G_{YN} t^{\gamma K}\,.
\label{eq-gdkyn-yn-M-IA+YN}
\end{eqnarray}
By inserting the Lippmann-Schwinger equation for $t_{YN}$, 
we get
\begin{eqnarray}
T_{IA+YN} &=& (1-V_{YN}\,G_{YN})^{-1}\,t^{\gamma K}\,.
\label{eq-gdkyn-yn-M-IA+YN-LS}
\end{eqnarray}
After solving the last equation in the partial wave decomposition, 
one obtains the $YN$ rescattering amplitude by subtraction of the
impulse term. 
The $KN$ rescattering (diagram (c)) is evaluated directly as
\begin{eqnarray}
T_{KN} &=& t_{KN}\, G_{KN}\, t^{\gamma K}\,,
\label{eq-gdkyn-kn}
\end{eqnarray}
where we use a separable potential of rank-1 for $V_{KN}$. 
The $\pi K$-process (diagram (d)) is calculated 
in the same way as in the $KN$-rescattering. 

\section{Results}
\label{sec-results}

The results are calculated by using 
the elementary operator of Mart and Bennhold\cite{MaB00} and 
the deuteron wave function of Bonn model\cite{MaH87}. 
\begin{figure}[htbp]
\centerline{\epsfxsize=8.6cm\epsfbox{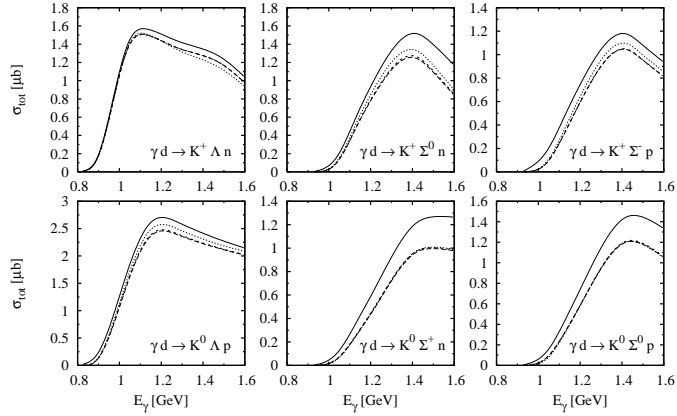}}   
\caption{Total cross section vs. photon energy $E_{\gamma}$.
\label{fig-gdkyn-tc}}
\end{figure}
\begin{figure}[htbp]
\centerline{\epsfxsize=8.6cm\epsfbox{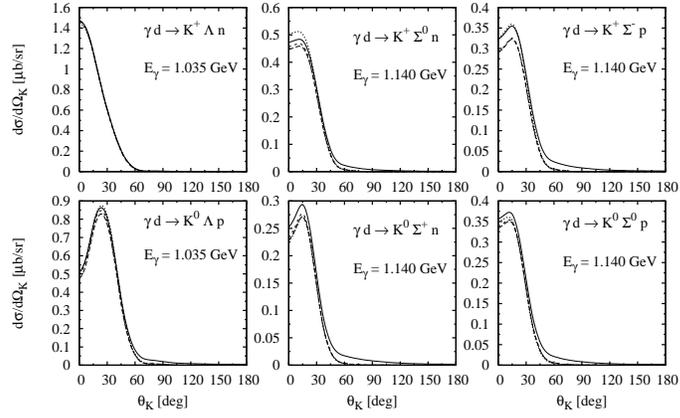}}   
\caption{Differential cross section vs. kaon angle $\theta_{K}$.
\label{fig-gdkyn-dc}}
\end{figure}
Fig.~\ref{fig-gdkyn-tc} shows the total cross section 
of $\gamma d \rightarrow KYN$ as function of $E_{\gamma}$. 
The $\pi K$-process (solid line) enhances the total cross section 
in all channels (dotted line). 
The next remarkable effect comes from $YN$ rescattering (short-dash line), 
while the effect of $KN$ rescattering (dash line) is negligible. 
Fig.~\ref{fig-gdkyn-dc} shows the differential cross section 
as function of $\theta_{K}$ calculated at different photon energies. 
$YN$ rescattering has remarkable effects at forward angles, 
while the $\pi K$-process at larger angles.
\begin{figure}[htbp]
\centerline{\epsfxsize=8cm\epsfbox{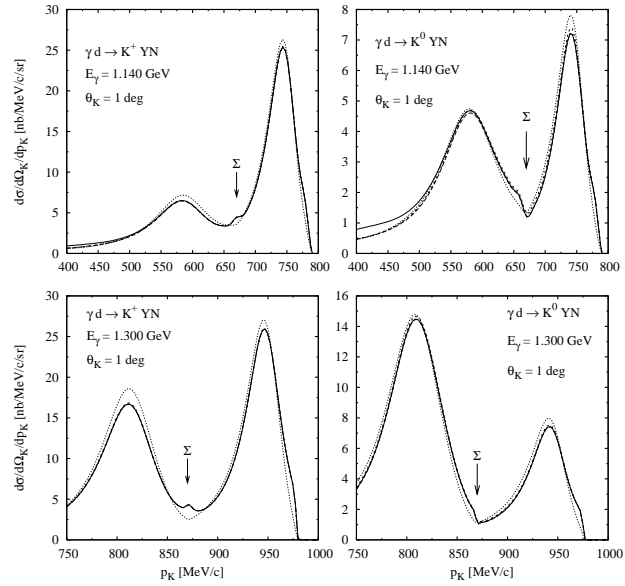}}   
\caption{Inclusive cross section vs. kaon momentum $p_{K}$.
\label{fig-gdkyn-inc}}
\end{figure}
The inclusive cross section as function of $p_{K}$ is shown 
in Fig.~\ref{fig-gdkyn-inc}. 
$YN$ rescattering shows remarkable effects 
at the threshold and peak region, while the effect of $\pi K$-process appears 
at smaller kaon momenta. Some enhancement, 
whose origin is from the $S$-matrix pole of $V_{YN}$\cite{MiY99}, 
is found at the $\Sigma$-threshold in the $\Lambda$-channels 
(indicated by arrows in the figure).  
\begin{figure}[htbp]
\centerline{\epsfxsize=9cm\epsfbox{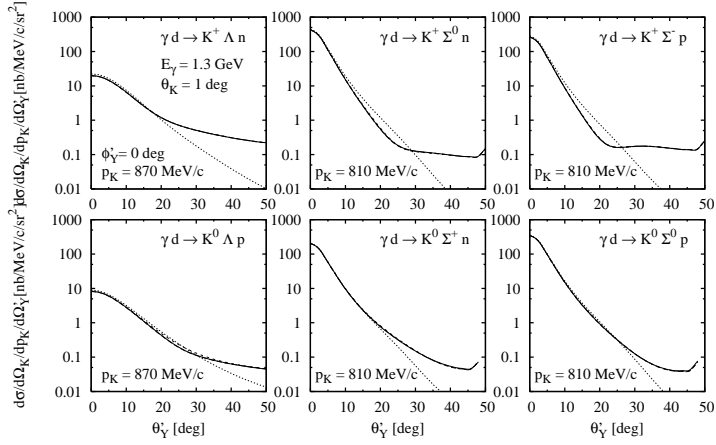}}   
\caption{Exclusive cross section vs. hyperon angle $\theta_{Y}'$.
\label{fig-gdkyn-exc}}
\end{figure}
Fig.~\ref{fig-gdkyn-exc} shows the exclusive cross section in the forward kaon angle 
at photon energy 1.3 GeV as function of $\theta_{Y}'$ 
measured relative to the direction of momentum transfer 
$\vec p_{\gamma}-\vec p_{K}$ in the deuteron rest frame. 
The $\Lambda$-channels are calculated at kaon momentum 870 MeV/c 
and $\Sigma$-channels at 810 MeV/c. $YN$ rescattering dominates the effect 
for all channels at larger $\theta_{Y}'$.

\section{Conclusions}
\label{sec-conclusions}

Kaon photoproduction on the deuteron is calculated by considering 
$YN$ and $KN$ rescattering and the two step process. 
A strong enhancement in the total cross section is found from the two-step process. 
$YN$ rescattering has remarkable effects in the inclusive and exclusive cross section, 
while the effect of $KN$ rescattering is negligible.

\section*{Acknowledgments}
AS would like to thank the Simulation Science Center, Okayama University of Science, 
Okayama for financial support and very kind hospitality.


\end{document}